\newcommand{\gcm}{{\rm g\,cm^{-2}}}
\newcommand{\Xf}{X_{\rm 1}}
\newcommand{\Xmax}{X_{\rm max}}
\newcommand{\Xmumax}{X^{\mu}_{\rm max}}

\documentclass[3p,times,article]{article}

\usepackage[figuresright]{rotating}

\begin{document}

\title{Constraints and measurements of hadronic interactions in extensive air showers with the Pierre Auger Observatory}
\date{}
\maketitle

\begin{center} 

L. Cazon$^1$ for the Pierre Auger Collaboration$^2$

\small{$^1$LIP, Av. Elias Garcia 14 - 1 1000-149 Lisbon,Portugal

$^2$Full author list: http://www.auger.org/archive/authors\_2013\_07.html
}
\end{center}

\begin{abstract}
The characteristics of extensive air showers are sensitive to the details of hadronic interactions at energies and in kinematic regions beyond those tested by human-made accelerators. Uncertainties on extrapolations of the hadronic interaction models in these regions hamper the interpretation of the ultra high energy cosmic ray data in terms of primary mass composition. We report on how the Pierre Auger Observatory is able to constrain the hadronic interaction models by measuring the muon content and muon production depth of air showers and also by measuring the proton-air cross section for particle production at a center-of-mass energy per nucleon of 57 TeV.

\end{abstract}

\section{Introduction}
\label{s:Introduction}

Interactions at a center of mass energy above those attained at the LHC are continuously happening in the upper layers of the Earth's atmosphere. They occur when ultra high energy cosmic rays (UHECR) collide with air nuclei,  being the highest energy so far recorded at $\sqrt{s}\sim$ 700 TeV, by Fly's Eye \cite{Elbert:1994zv}. In the decades to come, UHECR are the only way to explore such gigantic energies. Our current understanding of particle interactions at these energies relies on extrapolations made from experimental data collected in terrestrial human-made accelerators, which in addition are hampered by the difficulties of placing detectors in the most forward region. 

After the first cosmic ray interaction, thousands of secondaries interact again and cascade down to the Earth's surface, producing extensive air showers (EAS) of particles. The Pierre Auger Observatory detects those showers by sampling the EAS at ground with a surface detector array (SD), consisting of 1600 water Cherenkov stations separated by 1.5 km and spread over 3000 km$^2$. Fluorescence detectors (FD) collect light emitted by the passage of the charged particles of the shower through the air,  allowing the reconstruction of the longitudinal profile (LP) of the shower and a calorimetric measurement of its energy. Simultaneous detection by the SD and the FD is called hybrid detection and it has a dark night duty cycle of $\sim$15\% due to the FD. More details on the Observatory can be found in \cite{Abraham:2004dt,Allekotte:2007sf,Abraham:2009pm} and references therein.

The main goal of the Pierre Auger Observatory is to unveil the origin and nature of UHECR. A number of breakthroughs and some very important steps towards this goal have already happened: stringent photon limits have ruled out most top-down production mechanisms, favoring the acceleration scenarios in astrophysical sources. In addition, current neutrino limits have ruled out some exotic production models and are to reach the fluxes of cosmological origin \cite{Abreu:2011zze}, guaranteed if protons were the primaries. A flux suppression at $E=4 \times 10^{19}$ eV has been confirmed \cite{Abraham:2010mj}, being compatible with a GZK cut-off, but also with an energy exhaustion of the sources.  Arrival directions of the highest energy cosmic rays have been shown to be unevenly distributed in the sky, being correlated with the positions of nearby AGNs, which act as tracers of the extragalactic matter distribution \cite{Abraham:2007bb}\cite{Abreu:2010ab}.  The depth of the LP shower maximum is known to be sensitive to the primary mass composition, given that the deepest air showers occur for the smallest mass number $A$. As the energy of the shower increases, the shower gets larger and reaches its maximum development deeper in the atmosphere.   In general, a detailed simulation of the whole cascading process, accounting for all the multiparticle production details, is necessary to predict the position of the shower maximum as a function of mass and energy. Thus, mass interpretation can only be achieved by comparing the actual experimental readings with the predictions of full air shower simulations using the different high energy hadronic models.
 Results from the Pierre Auger Observatory show a composition which steadily becomes heavier with energy when compared to the latest available models \cite{Abraham:2010yv}.  The number of muons at the ground is also sensitive to the mass of the primaries \cite{Kampert:2012mx}, but it is also hampered by the ambiguity of the predictions of the high energy interaction models.  The phase space of shower observables occupied by different primary masses  often overlaps with that of the different model predictions. Disentangling one from the other is of utmost importance and is one of the most compelling challenges in UHECR physics.

In this paper we focus on the Pierre Auger measurements relevant to constrain our knowledge of high energy physics. In section 2 
a measurement of proton-air cross section is presented and in section 3, measurements are presented related to the muon production in extensive air showers, namely, different measurements of the muon number at the ground are described in subsections 3.1, 3.2, and 3.3, 
 and the longitudinal production profile is described in subsection 3.4. 

\section{Measurement of the proton-air cross section}
\label{sec1}
The depth at which the parent cosmic ray interacts, $\Xf$, follows an exponential distribution $\propto \exp{\left(-\frac{\Xf}{\lambda}\right)}$ where $\lambda$ is inversely proportional to the p-air cross section, $\sigma^{prod}_{p-air}$,  that accounts for all interactions which produce particles, and thus contribute to the air shower development; it implicitly also includes diffractive interactions.  The depth required for the shower to fully develop is $\Delta X$, being the tail of the $\Xmax$-distribution of proton showers directly related to the distribution of the first interaction point $\Xf$ through $\Xmax=\Xf+\Delta X$. Thus,
\begin{equation}
\frac{dN}{d\Xmax}=N \exp{\left(-\frac{\Xmax}{\Lambda_\eta}\right)}
\end{equation}
where  $\eta$ represents the fraction of the most deeply penetrating air showers used. Thus, $\eta$ is a key
parameter: a small value enhances the proton fraction, but
reduces the number of events available for the analysis. We
have chosen $\eta=0.2$ so that, for helium-fractions up to
25\%, biases introduced by the possible presence of helium
and heavier nuclei do not exceed the level of the statistical
uncertainty.
 
\begin{figure}[h]
\begin{center}
\begin{minipage}{7cm}
\includegraphics[width=7cm]{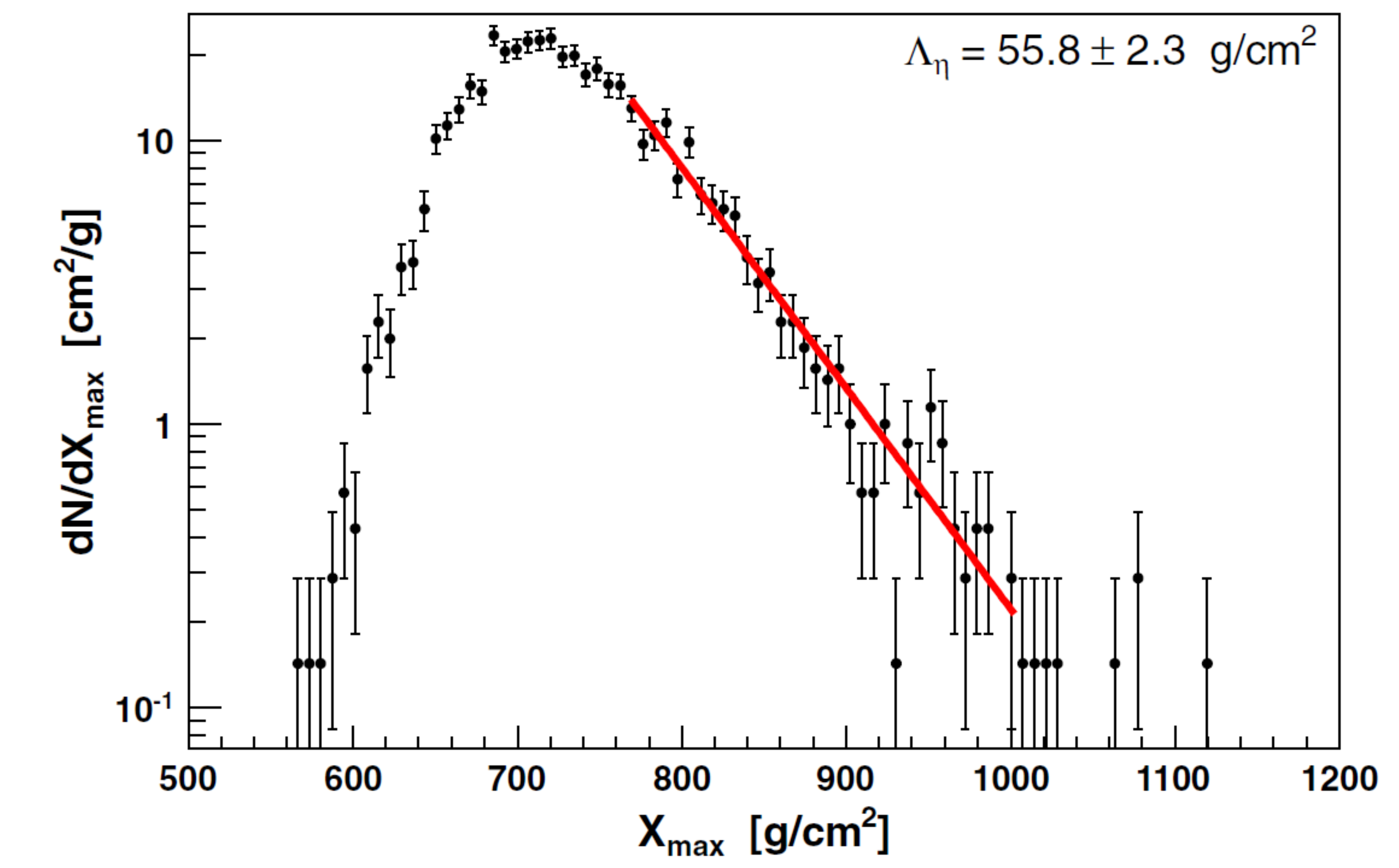}
\end{minipage}\hspace{2pc}%
\begin{minipage}{7cm}
\includegraphics[width=7cm]{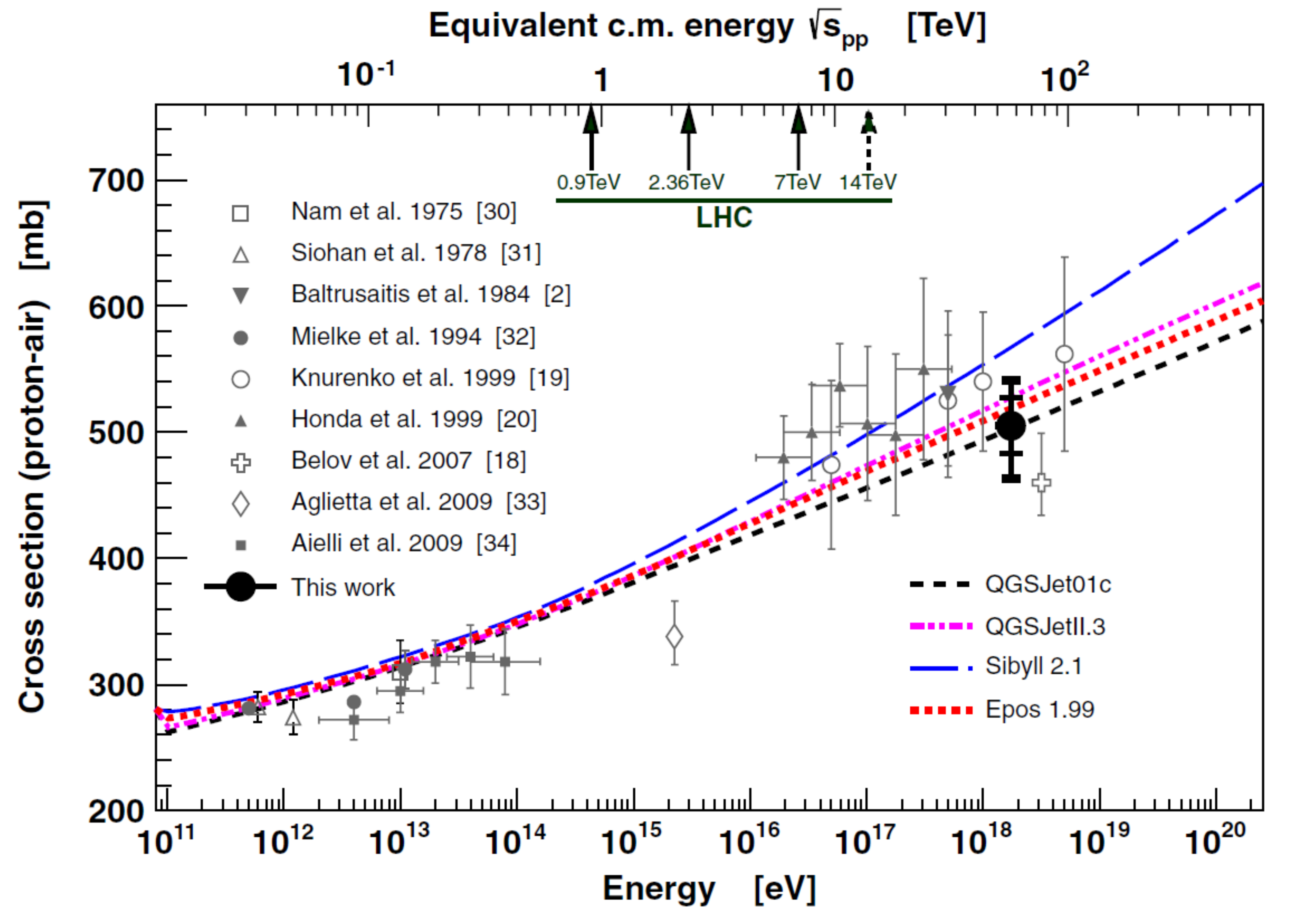}
\end{minipage} 
 \caption[]{Left panel:Unbinned likelihood fit to the $\Xmax$-distribution to obtain $\Lambda_\eta$ (thick line). Right panel: Resulting
p-air cross section compared to other measurements and different model predictions. The inner error bars are statistical, while the outer include systematic uncertainties for a helium fraction of 25\% and 10 mb for the systematic uncertainty attributed to the
fraction of photons. See \cite{Collaboration:2012wt} for details.}
 \label{Crosssection}
\end{center} 
\end{figure}

Figure  \ref{Crosssection} displays the $\Xmax$ distribution for selected events, resulting in a value 
\begin{equation}
\Lambda_\eta=55.8 \pm 2.3{\rm(stat)} \pm 1.6 {\rm(sys)} \, \, {\rm g\, cm^{-2}}. 
\end{equation}
The average energy  is $10^{18.24}$ eV which corresponds to a center-of-mass
energy of $\sqrt{s}=57$ TeV in proton-proton collisions.

The hadronic cross sections of the different high energy interaction models were multiplied by an energy-dependent factor
\begin{equation}
f(E,f_{19})=1+(f_{19}-1) \frac{\log(E/10^{15}{\rm eV})}{\log 10^{19}/10^{15}}
\end{equation}
to produce different predictions of the slope, $\Lambda_\eta^{MC}$. This allows us to directly relate the measured $\Lambda_\eta$ to the corresponding $\sigma^{prod}_{p-air}$ for a given model.  

After averaging the four values
of the cross section obtained with the different available hadronic interaction
models we obtain:
\begin{equation}
\sigma^{prod}_{p-{air}}=505\pm 22{\rm(stat)}^{+28}_{-36} {\rm(sys)} \, \, {\rm mb}
\end{equation}
at a center-of-mass energy of $57\pm 0.3{\rm(stat)}\pm 6{\rm(sys)}$ TeV.

The results are
presented assuming a maximum contamination of 25\% of
helium nuclei in the light cosmic-ray mass component. The lack
of knowledge of the helium component is the largest
source of systematic uncertainty. However, for helium
fractions up to 25\% the induced bias remains below 6\%. More details of this analysis can be found in \cite{Collaboration:2012wt}.

\section{Muon Production in Air Showers}
\label{sec2}
The hadronic cascade is the main engine that drives the development of nuclei-induced EAS. Approximately 80\% of the particles produced in  high energy collisions are pions, of which approximately 1/3 are neutral pions. They rapidly decay into photons, feeding the electromagnetic (EM) cascade. After only three hadronic generations, $\sim$ 50\% of the energy is already transferred to the EM cascade \cite{Cazon:2013dm}, causing it to rapidly decouple from the hadronic cascade. After each interaction, the other 2/3 of the energy keeps feeding the hadronic cascade through charged pions every generation, until they decay into muons at a few tens of meters from the shower axis \cite{Cazon:2012ti}, and then leaving the central region of the shower. Muons travel to the ground almost in straight lines, as Coulomb scattering is less important than for electrons. They act as true messengers from the hadronic skeleton of the shower, and allow us to peer into details of the hadronic physics at the core of the EAS.

Despite of the Pierre Auger Observatory being an experiment not having been originally designed to separately measure air shower components, we have developed techniques that allow us to assess the muon and the electromagnetic contributions, either by analyzing the time structure of the signals in the SD stations, or by analyzing different regions in terms of core distance and zenith angle of the showers, where the muon component is dominant.

Whenever a charged particle of sufficient energy passes through water of a SD station, it produces Cherenkov photons. After a few reflections into the wall material, made of Tyvek, their distribution is isotropized. Their concentration is sampled by the FADCs from three photomultiplier tubes viewing the water volume, before being absorbed by the water after $\sim$ 100 ns.  The observed signal is basically proportional to the track-length that the particle traverses in water, and therefore there is not a basic difference between signals produced by muons compared to those produced by electrons or positrons. 

 Given that electrons are far more numerous than muons and that they typically cascade down inside the water, while muons typically traverse the water without interactions but energy loss, the typical track-lengths of electrons in water are shorter. As a consequence, the EM component of the shower produces a signal distribution in time which is smoother than the muonic one, which is spiky and can be discriminated under some conditions. Note nevertheless that converting photons might also produce signals with very similar characteristics to those of muons.

In addition, the relative richness of muons compared to EM particles in EAS increases with the distance to the shower core, and also with the zenith angle of the EAS.

The Pierre Auger Collaboration has developed different techniques to assess the muon content of EAS under different conditions, namely, 1) analysis of the muon fraction through the temporal structure of the SD signals in vertical showers, 2) measurement of a hadronic scale factor by analyzing the signal size in vertical hybrid events, and finally 3) analysis of the muon content in inclined hybrid showers. 

\subsection{Muon fraction through the temporal structure of the SD signals}
\label{subsec1}
Given that the number of muons at 1000 m from the core scales nearly linearly with the energy of the shower, the fraction of the signal attributed to muons to the total signal, $f_{\mu}=S_{\mu}/S$ is insensitive to the systematic uncertainty of the energy, which is 14\% (\cite{ValerioICRC2013}).
Two different methods were used to assess $f_{\mu}$: a multivariate method, and a smoothing method.

The basic idea of the multivariate method is to combine muon-content sensitive characteristics of the FADC signal to reconstruct $f_{\mu}$ using:
\begin{equation}
\hat {f}_{\mu} = a+b \hat{\theta} +c f^2_{0.5}+d \hat{\theta} P_0+ e \hat{r}
\end{equation}
where $\hat{\theta}$ is the reconstructed zenith angle of the shower
and $\hat{r}$ is the distance of the detector from the reconstructed
shower axis. $f_{0.5}$ is the portion of the signal in FADC bins
larger than 0.5 Vertical Equivalent Muons (VEM),  and $P_0$ is the normalized zero-frequency component
of the power spectrum \cite{BalaszICRC2013}.

Both $f_{0.5}$ and $P_0$ are sensitive
to large relative fluctuations and short signals, which
are the signatures of high muon content. 
We estimate the fit parameters $(a,b,c,d,e)$ using simulations described in \cite{BalaszICRC2013}.

\begin{figure}[h]
\begin{center}
\includegraphics[width=10cm]{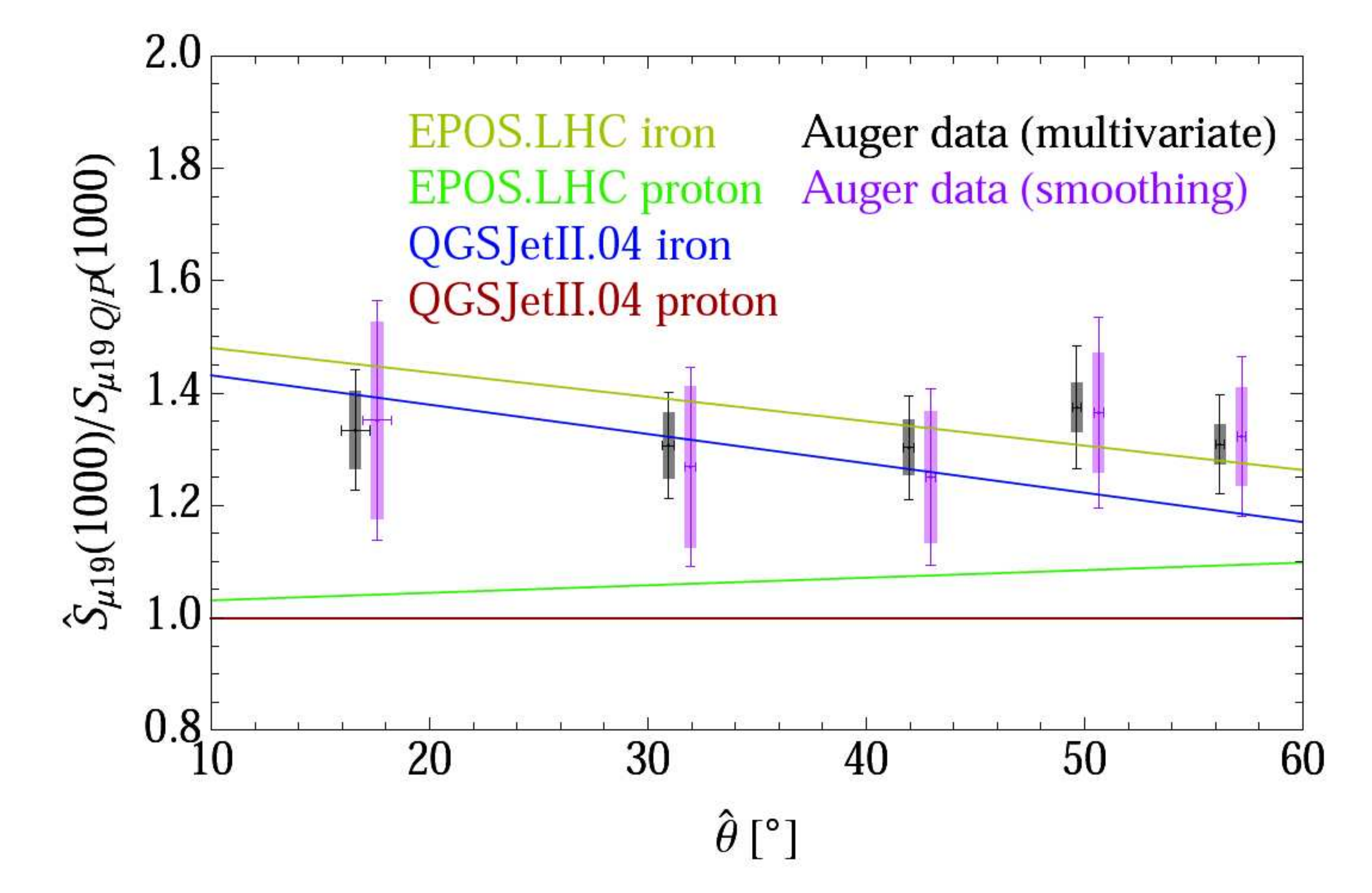}
 \caption[]{The measured muon signal rescaling to $E = 10^{19}$ eV
and at 1000 m from the shower axis vs. zenith angle, with respect
to QGSJetII-04 proton simulations as a baseline. The rectangles represent the
systematic uncertainties, and the error bars represent the statistical
uncertainties added to the systematic uncertainties. See \cite{BalaszICRC2013} for details.}
 \label{f191}
\end{center} 
\end{figure}

The smoothing method is a low-pass filter, which was run
a few times on the signal to gradually separate the low-frequency EM component from the high-frequency one which is attributed to muons. Firstly, the signal is smoothed  by a moving average of size $L$ over the FADC. 
The window size $L$ was adjusted using simulations to follow the low frequencies corresponding to the EM signal at
large angles, while narrower windows are needed to extract
it in vertical showers, where the EM component is more
similar to the muonic signal, resulting in  $L = 7.83+0.09\theta/$deg.
The procedure was repeated four times, re-smoothing each time the output of the previous iteration. The final muonic signal is the sum of the non-smooth
positive differences at each step. 

The muon signal can be retrieved by multiplying $\hat{f}_{\mu}$ by the total signal. The results respect
to QGSJetII-04 \cite{Ostapchenko:2010vb} proton simulations are shown in Fig. \ref{f191} for $E = 10^{19}$ eV and $r = 1000$ m, with a value $\sim 1.3-1.4$
as a function of the zenith angle. Good agreement is found between the two analysis methods.
The model predictions for proton- and iron-induced
showers bracket the measurements within the
systematic uncertainties. More details on this analysis can be found in \cite{BalaszICRC2013}.

\subsubsection{Signal size of vertical hybrid events}
\label{subsec2}
The ground signal of simulated
showers with longitudinal profiles matching those of detected showers was analyzed.
The data used for this study 
were narrowed down to the energy bin 10$^{18.8} < E < 10^{19.2}$ eV, sufficient to have adequate statistics while being narrow enough that
the primary cosmic ray mass composition does not evolve
significantly.

\begin{figure}[h]
\begin{center}
\includegraphics[width=10cm]{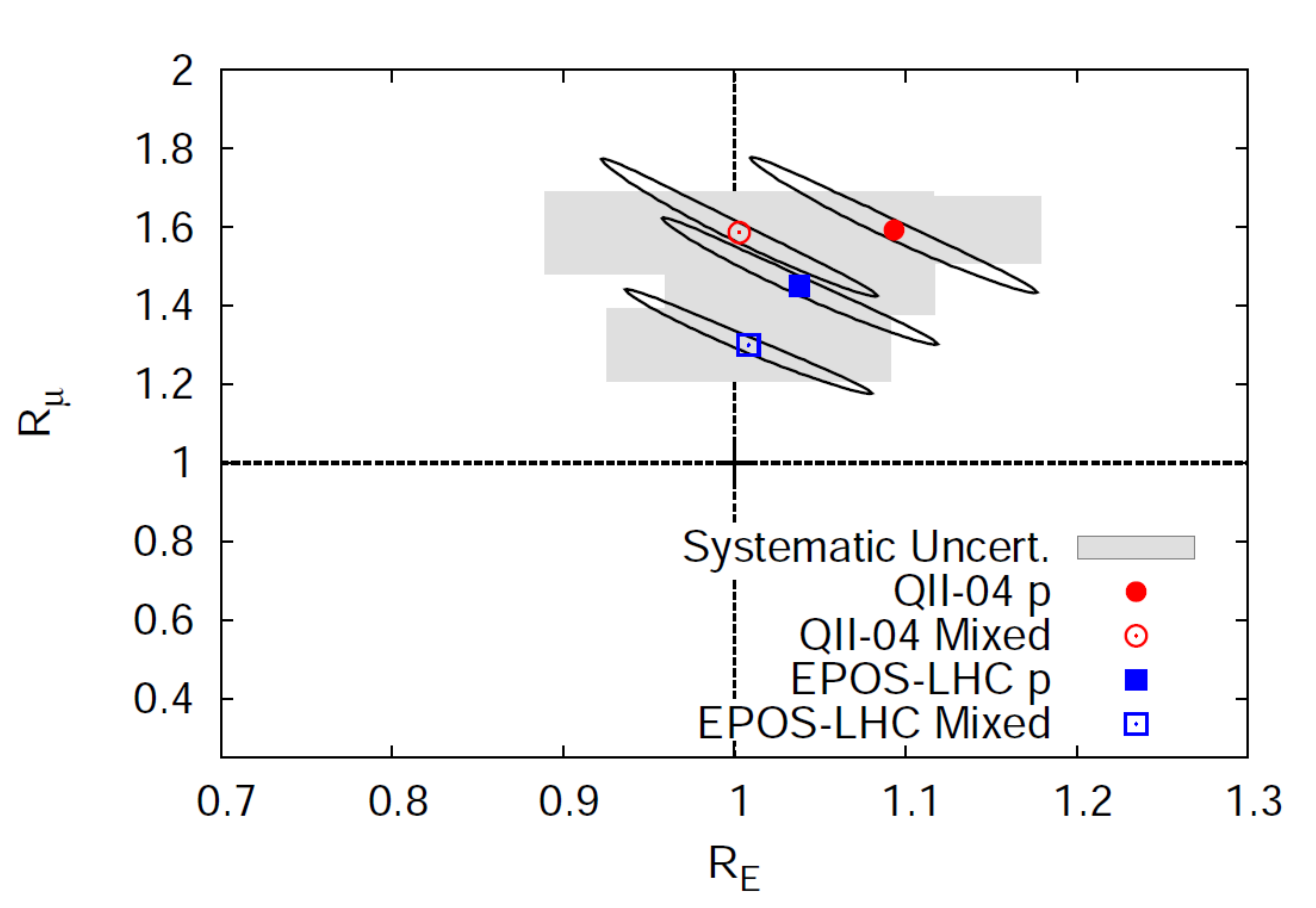}
 \caption[]{Value of the hadronic rescaling parameter $R_{\mu}$ and the energy rescaling parameter $R_{E}$ for Auger hybrid data at 10 EeV. See \cite{GlennysICRC2013} for more details.}
 \label{RH1}
\end{center} 
\end{figure}

Each event was compared with the results obtained from simulations using two different hadronic models (QGSJetII-04 
and EPOS-LHC \cite{Pierog:2013ria}) and for four different primary masses (proton, helium, nitrogen, and iron), for all of the
events in the dataset.

To explore the potential sources of the muon count discrepancy between measurements and model expectations, the
ground signal was modified in the simulated events to fit the
ground signal in the data. Two rescaling factors were introduced:
$R_{\rm{E}}$ and $R_{\mu}$. $R_{\rm{E}}$ acts as a rescaling of the energy of
the primary cosmic ray, affecting the total ground signal. $R_{\mu}$ acts as a rescaling
factor of the contribution to the ground signal
of inherently hadronic origin.
 $R_{\rm{E}}$ and $R_{\mu}$ are then fitted to minimize the discrepancy between the ensemble of
observed and simulated signals at ground, which can also reproduce the observed $\Xmax$-distribution, and is labeled as ``mixed'' in Fig. \ref{RH1}.
The observed hadronic signal is a factor 1.3 to 1.6 larger than
predicted using the hadronic interaction models
tuned to fit LHC and lower energy accelerator data. None of the tested models calls for an  energy
rescaling. More details of this analysis can be found in \cite{GlennysICRC2013}.

\subsubsection{Signal size of inclined hybrid events}
\label{subsec3}
After the arrival direction ($\theta$,$\phi$) of the cosmic ray is
determined from the relative arrival times of the shower
front,
the shower size parameter $N_{19}$ is defined through the following relation:
\begin{equation}
\rho_{\mu} = N_{19} \, \rho_{\mu,19}(x,y,\theta,\phi), \label{eq_HAS}
\end{equation}
where $\rho_{\mu}$ is the model prediction for the muon density
at the ground used to fit the signals recorded at the detectors.
$\rho_{\mu,19}$ is a reference profile corresponding to the inferred
arrival direction, obtained as a parametrization \cite{Dembinski:2009jc} of
the muon density at ground for proton showers of 10$^{19}$ eV,
simulated using 
the QGSJetII-03
 interaction model. 
$N_{19}$ is sensitive to the cosmic-ray energy and nuclear mass composition. The quantity $R_{\mu}$ ($R_{\mu}\simeq N_{19}$) was introduced to account for the difference between the real number of muons, given by the  integral of the distribution of muons at the ground, and the estimate obtained by the fitting procedure of eq. \ref{eq_HAS}. The difference between $N_{19}$ and $R_{\mu}$ is less than 5\%.

\begin{figure}[h]
\begin{center}
\includegraphics[width=10cm]{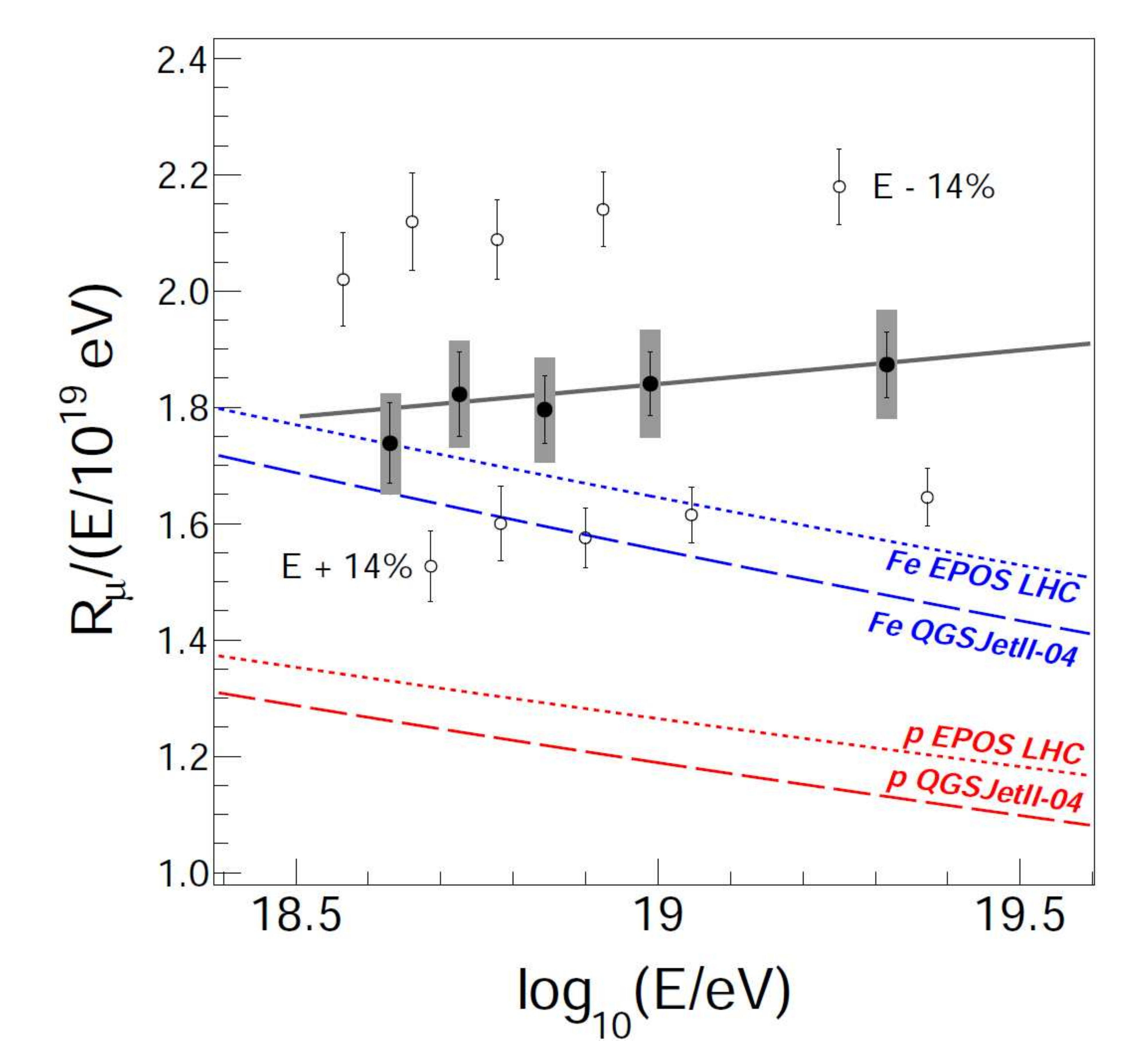}
 \caption[]{Average value of $R_{\mu}$/($E_{FD}/10^{19}$ eV) as a function
of shower energy. The
gray thick error bars indicate the systematic uncertainty.
 Theoretical curves for proton and
iron showers simulated with QGSJetII-04 and EPOS LHC
are shown for comparison. Open circles indicate the result
if the FD energy scale is varied by its systematic uncertainty. See \cite{InesICRC2013} for more details.
}
 \label{HAS1}
\end{center} 
\end{figure}

The averaged scaled quantity 
$R_{\mu}$ / ($E_{FD}$ / 10$^{19}$ eV) is shown in Fig. \ref{HAS1} 
divided in five energy bins containing roughly equal statistics.
The measurement of $R_{\mu}$ / ($E_{FD}$ / 10$^{19}$ eV) is dominated by systematic uncertainties in the energy
scale (shown as open circles in the figure). The measured number of muons between 4 $\times$ 10$^{18}$ eV and 2 $\times$ 10$^{19}$ eV is
marginally comparable to predictions for iron showers simulated either
with QGSJetII-04 or EPOS-LHC if we allow the FD energy scale to increase by its systematic uncertainty
of about 14\% (\cite{ValerioICRC2013}).

Given that the observed distribution of the depth of
shower maximum between $4\times10^{18}$ eV and $2\times10^{19}$ eV is
not compatible with an iron dominated composition, 
we conclude that the observed number of muons is not
well reproduced by the shower simulations. More details of this analysis can be found in \cite{InesICRC2013}.

\subsection{Muon production depth}
\label{subsec3}
The distribution of muon arrival times to the ground is closely related to the distribution of their production depths.
To a first approximation, there is a one-to-one map between the time elapsed between the arrival time of a hypothetical shower front plane, traveling at the speed of 
light, and the arrival time of the muons whose trajectories are not parallel to the shower axis: $ct_g=\sqrt{r^2+(z-\Delta)^2}-(r-\Delta)^2$,
where $r$ is the distance to the shower core in the perpendicular plane, $z$ is the distance from the ground to the production point, and $\Delta$ is the $z$-coordinate of the observation point. Both $\Delta$ and $z$ are measured along the shower axis.

The second most important source of delay is the subluminal velocities of the muons, due to their finite energy \cite{Cazon:2012ti}. The so called kinematic time is a second order correction to the total arrival time delay, ($< \sim$ 10\% above 1000 m from the core), that decreases as $r$ increases. Its average $\langle ct_\epsilon \rangle$ is calculated from an analytic model for the energy spectrum of muons \cite{Cazon:2003ar}.

The production distance $z$ is approximated as
\begin{equation}
z\simeq\frac{1}{2}\frac{r^{2}}{ct-\langle ct_\epsilon \rangle} + \Delta
\end{equation}
which is later  transformed into a production depth using the density profiles provided by the instruments dedicated to monitor the atmosphere above the Auger Observatory.

\begin{figure}[h]
\begin{center}
\includegraphics[width=10cm]{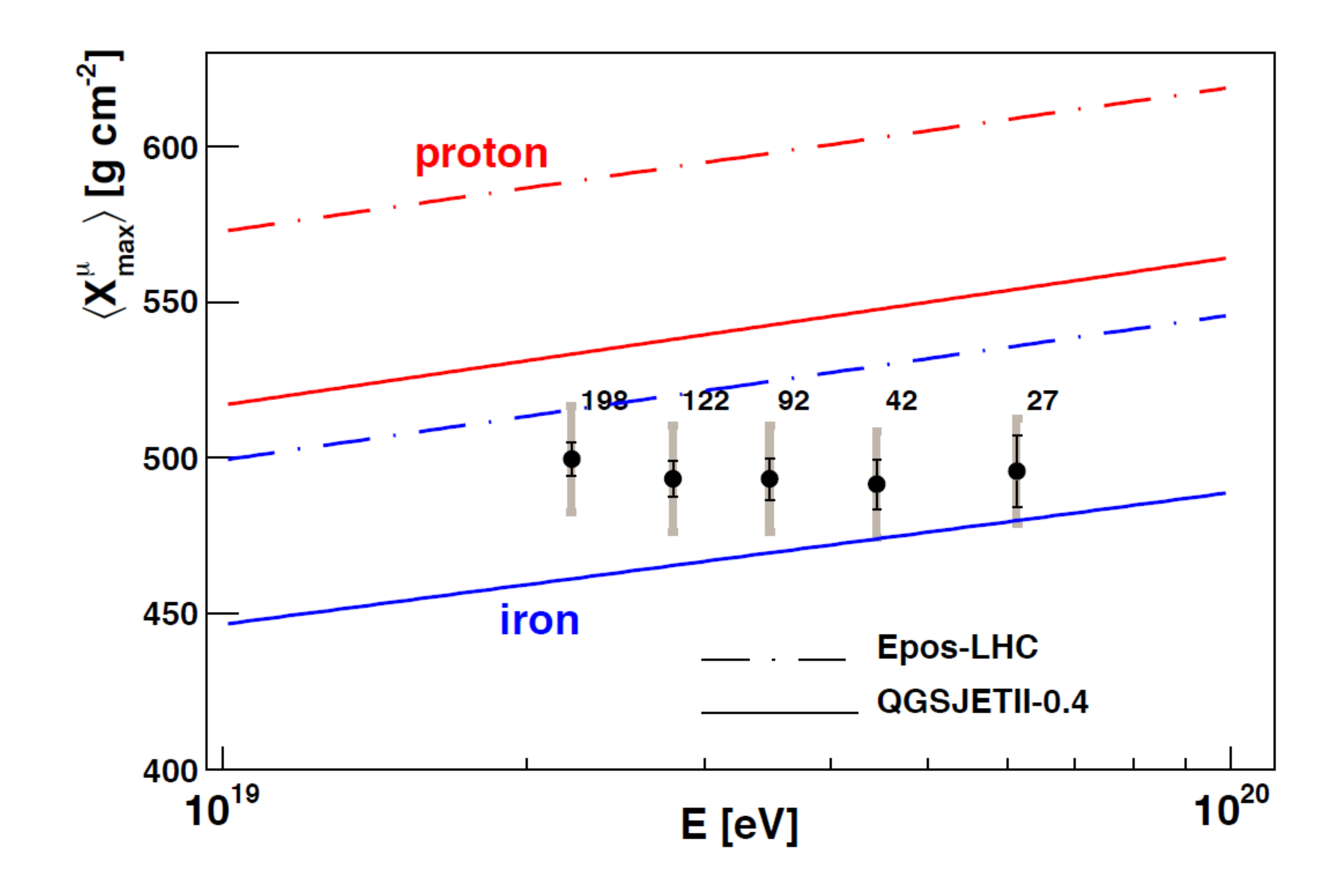}
 \caption[]{$\langle \Xmumax \rangle$ as a function of energy. The prediction of
different hadronic models for proton and iron primaries are shown.
Numbers indicate the number of events in each energy
bin and the gray rectangles represent the systematic
uncertainty. See \cite{DiegoICRC2013} for details.}
 \label{MPD1}
\end{center} 
\end{figure}

The data set used in this analysis comprises the events
recorded in the angular range from 55$^{\circ}$ to 65$^{\circ}$.
The evolution of the measured average maximum of the muon production depth distribution $\langle \Xmumax \rangle$ as a function of
$\log_{10}(E/\rm{eV})$ is shown in Fig. \ref{MPD1}. The uncertainties represent the standard error on the mean, whereas the gray bars represents the systematic uncertainty, which amounts to 17 $\gcm$. Fig. \ref{MPD1} also displays QGSJetII-04 and EPOS-LHC predictions for both proton and iron primaries. Both models have the same muonic
elongation rate but with considerable differences in the absolute value of $\langle \Xmumax \rangle$. More details of this analysis can be found in \cite{DiegoICRC2013}.

\begin{figure}[h]
\begin{center}
\includegraphics[width=15cm]{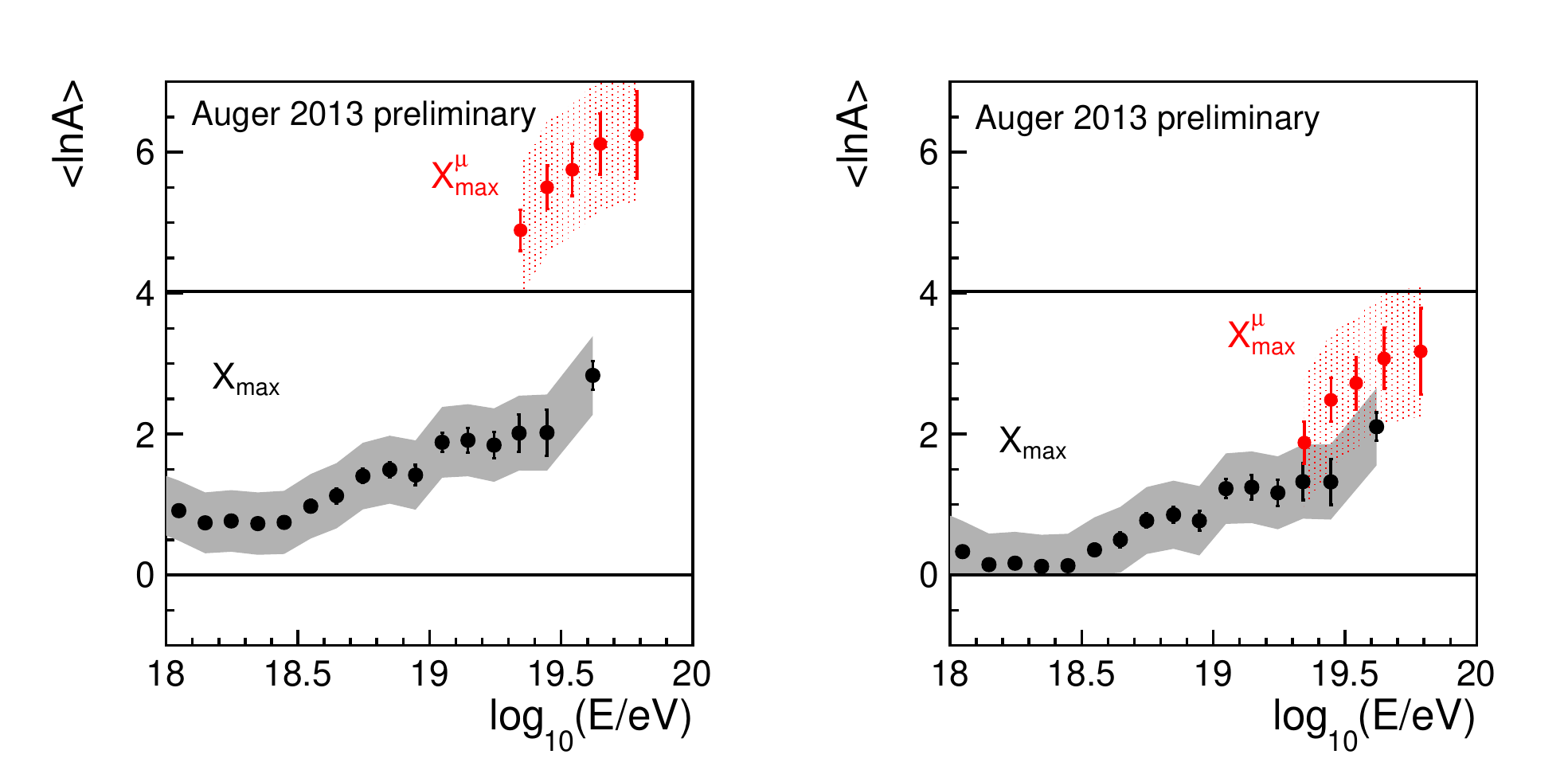}
 \caption[]{Conversion of the $\langle \Xmax \rangle$ and $\langle \Xmumax \rangle$  into $\langle lnA \rangle$ using two different hadronic interaction models, EPOS-LHC (left) and QGSJetII-04 (right). The solid horizontal lines are the reference values from proton (botton) and iron (top) primaries. The bands represent the systematic uncertainties. See \cite{AntoineICRC2013} for details.}
 \label{lnA}
\end{center} 
\end{figure}

It is possible to linearly convert $\langle \Xmax \rangle$ and $\langle \Xmumax \rangle$ into the mean logarithmic mass of the primary, $\langle lnA \rangle$, for a given high-energy interaction model. A mismatch of this conversion would 
necessarily imply that such model is unable to consistently predict for the same primary the values of $\langle \Xmax \rangle$ and the values of $\langle \Xmumax \rangle$ \cite{AntoineICRC2013}.

For the EPOS-LHC model, this conversion procedure results into incompatible $\langle \ln A \rangle$ values, and the mass conversion of $\langle\Xmumax \rangle$ resulting in $\langle \ln A \rangle \sim 5$, a value that corresponds to a nuclei which is much heavier than iron $lnA\simeq 4$ well beyond the systematic uncertainties. 
The procedure using the second model, QGSJetII-04, yields self-consistent results in this respect. 
It can be observed from Fig. \ref{MPD1} that EPOS-LHC predicts reference lines for proton and iron primaries much deeper than older versions and other models. 
Paradoxically, EPOS-LHC is claimed \cite{EPOSclaim} to better represent the rapidity gap distributions of the new LHC p-p data, when compared to QGSJetII.04. 
This fact could be symptomatic of further misadjustments that partially compensate each other in the rest of the models.
This analysis shows that UHECR shower studies provide handles to constrain high-energy interaction models.

\section{Conclusions}

The Pierre Auger Observatory has unique capabilities to make measurements of the highest energy hadronic interactions and to constraint the models that attempt to describe them. The proton-air cross section has been measured at $\sqrt{s}=57$ TeV, which is compatible with extrapolations made using high-energy interaction models.
Secondly, despite not being originally designed for such a purpose,
and by means of indirect methods, the Pierre Auger Observatory is able to measure both the longitudinal development of the muon production and the muon content at the ground under various conditions.  The observed muon content in 10 EeV air showers ($\sqrt{s}$ = 137 TeV) is a factor $\sim 1.3-1.6$ larger than
predicted using the latest hadronic interaction models tuned to LHC and lower energy accelerator data. The best model for describing such data is EPOS-LHC, which in turn fails to describe the maximum of the muon production depth distribution by a large factor. Paradoxically, EPOS-LHC is claimed to better describe some of the multiparticle production observables compared to QGSJetII.04, namely the rapidity gap distributions. This could possibly mean the existence of further phenomena  not yet accounted for, at least with enough accuracy within the models.

\end{document}